\documentstyle[prl,aps,multicol]{revtex}

\draft

\author{M.~A.~Nielsen}
\title{Characterizing mixing and measurement in quantum mechanics}
\address{Physics MC 12-33, California Institute of Technology,
  Pasadena CA 91125, USA \\
  Center for Quantum Computer Technology, 
  University of Queensland 4072, Australia}
\date{\today}

\begin{document}

\pagestyle{plain}
\pagenumbering{arabic}

\maketitle

\begin{abstract}
  What fundamental constraints characterize the relationship between a
  mixture $\rho = \sum_i p_i \rho_i$ of quantum states, the states
  $\rho_i$ being mixed, and the probabilities $p_i$?  What fundamental
  constraints characterize the relationship between prior and
  posterior states in a quantum measurement?  In this paper we show
  that there are many surprisingly strong constraints on these mixing
  and measurement processes that can be expressed simply in terms of
  the eigenvalues of the quantum states involved.  These constraints
  capture in a succinct fashion what it means to say that a quantum
  measurement {\em acquires} information about the system being
  measured, and considerably simplify the proofs of many results about
  entanglement transformation.
\end{abstract}


\begin{multicols}{2}[]
\narrowtext

\section{Introduction}

%
%
Quantum mechanics harbours a rich structure whose investigation and
explication is the goal of quantum information
science\cite{Nielsen00a,Preskill98c}.  At present only a limited
understanding of the fundamental static and dynamic properties of
quantum information has been obtained, and many major problems remain
open.  In particular, we would like a detailed ontology and
quantitative methods of description for the different types of
information and dynamical processes possible within quantum mechanics.
An example of the pursuit of these goals along a specific line of
thought has been the partial development of a theory of entangled
quantum states; see for example the work in
\cite{Bennett96a,Nielsen99a,Horodecki96f,Vedral98a,Vidal99a,Vidal00b,Jonathan99a,Bennett99b,Terhal00a,Dur00a}.

%
%
The purpose of the present paper is to pose and partially solve two
fundamental problems about the static and dynamic properties of
quantum information.  The first of these problems is to characterize
the process of {\em mixing} quantum states.  More precisely, if $\rho
= \sum_i p_i \rho_i$ is a mixture of quantum states $\rho_i$ with
probabilities $p_i$, what constraints relate the properties of $\rho$
to the probability distribution $p_i$ and the quantum states $\rho_i$?
The second problem is to characterize the relationship between the
prior and posterior states in a quantum measurement.  The result of
our investigations is a set of two {\em static constaints} on mixtures
of quantum states, two {\em dynamic constraints} on the quantum
measurement process, and two partial {\em converse} results, one to
the static constraints, and the other to the dynamic constraints. The
statement of each of these results is rather easily understood, so we
review the statements now, before proceeding to the proofs and
consequences in the main body of the paper.

%
%
Suppose we mix a set of quantum states $\rho_i$ according to the
probability distribution $p_i$.  Then we will show that this mixing
process must satisfy the constraint equations:
\begin{eqnarray} \label{eq:maj_1_intro}
  \lambda\left(\sum_i p_i \rho_i \right) & \prec & \sum_i p_i \lambda(\rho_i) \\
  \label{eq:maj_2_intro}
  \bigoplus_i p_i \lambda(\rho_i) & \prec & \lambda\left( \sum_i p_i
  \rho_i \right).
\end{eqnarray}
In these equations the notation $\oplus$ denotes a direct sum of
vectors, $\lambda(X)$ denotes the vector of eigenvalues of the matrix
$X$ arranged so the components appear in non-increasing order, and the
relation ``$\prec$'' is the {\em majorization} relation\footnote{Note
  that the vectors on the left and right hand sides
  in~(\ref{eq:maj_2_intro}) may be of different dimension; in such
  cases we extend whichever vector is of lesser dimension by padding
  it with zero entries, to enable comparison using the majorization
  relation.}.  As an example of the notation used
in~(\ref{eq:maj_2_intro}), suppose $p_1 = 1/3, p_2 = 2/3, \rho_1 =
\mbox{diag}(3/4,1/4)$ and $\rho_2 = \mbox{diag}(1/5,4/5)$.  Then
Equation~(\ref{eq:maj_2_intro}) becomes
\begin{eqnarray}
  \frac{1}{3} \left[ \begin{array}{c} \frac{3}{4} \\ \frac{1}{4}
    \end{array} \right]
  \bigoplus \frac{2}{3} \left[ \begin{array}{c} \frac{4}{5} \\
  \frac{1}{5} \end{array} \right] \prec
  \lambda\left( \frac{1}{3} \left[ \begin{array}{cc}
        \frac{3}{4} & 0 \\ 0 & \frac{1}{4} \end{array} \right]
    + \frac{2}{3} \left[ \begin{array}{cc} \frac{1}{5} & 0 \\
        0 & \frac{4}{5} \end{array} \right] \right),
\end{eqnarray}
which is equivalent to
\begin{eqnarray}
  \left[ \begin{array}{c} \frac{1}{4} \\ \frac{1}{12} \\
      \frac{8}{15} \\ \frac{2}{15} \end{array} \right]
  \prec \left[ \begin{array}{c} \frac{37}{60} \\ \frac{23}{60} \\ 0 \\ 0
    \end{array} \right].
\end{eqnarray}
A formal definition of majorization appears in
Subsection~\ref{subsec:majorize}, however for now the essential
intuition to grasp is that the relation $x \prec y$ means that the
vector $x$ is more ``mixed'' (or ``disordered'') than $y$.  Thus,
Equation~(\ref{eq:maj_1_intro}) captures the intuition that $\sum_i
p_i \rho_i$ is more mixed, on average, than the states $\rho_i$
appearing in the ensemble.  The intuition
behind~(\ref{eq:maj_2_intro}) is a little more complex.  Imagine that
we prepare the state $\rho$ by randomly choosing a value for $i$
according to the probability distribution $p_i$, and then preparing
the corresponding state $\rho_i$.  Our quantum state, including a
description of $i$, may be written as $\sum_i p_i |i\rangle \langle i|
\otimes \rho_i$.  We then ``throw away'' the state $|i\rangle$
representing our random choice of $i$, leaving only the state $\sum_i
p_i \rho_i$.  The relation~(\ref{eq:maj_2_intro}) expresses the fact
that when we throw away $i$, the state of the quantum system becomes
less disordered.

%
%
Suppose we perform a measurement on a quantum mechanical system
initially in the state $\rho$, obtaining measurement result $i$ with
probability $p_i$, and corresponding posterior state $\rho_i'$.  What
constraints are placed on the relationship between $\rho$, $p_i$ and
$\rho_i'$?  We will show that the following two {\em dynamic
  constraints} must be satisifed:
\begin{eqnarray} \label{eq:msment_1}
\lambda(\rho) & \prec & \sum_i p_i \lambda\left( \rho_i'
\right) \\ \label{eq:msment_2}
\bigoplus_i p_i \lambda\left( \rho_i' \right) & \prec &
\lambda\left( \rho \right).
\end{eqnarray}
The intuition behind~(\ref{eq:msment_1}) is that quantum measurements
acquire information about the state of the system being measured, and
thus after measurement the state of the system is less mixed, on
average, than before.  The intuition behind~(\ref{eq:msment_2}) is a
little more complex, but can be understood using Zurek's
approach\cite{Zurek91a} to decoherence and quantum measurement.
Recall that in this approach a measurement involves three systems: the
system being measured, which starts in the state $\rho$, and ends in
the state $\rho_i'$; a measuring device, which starts in some standard
state, and finishes in a ``pointer state'' $|i\rangle$ recording the
result of the measurement, and an environment which ``decoheres'' the
measuring device, ensuring that it behaves in an essentially classical
fashion.  The system and measuring device interact unitarily during
the measurement, ensuring that there is no change in the amount of
disorder present in the system.  The subsequent environmental
decoherence process can also be thought of as a type of measurement,
in which the different outcomes are averaged over.  In this view, the
environment continually measures the state of the measuring apparatus,
resulting in a final state $\sum_i p_i |i\rangle \langle i| \otimes
\rho_i'$ for the measuring apparatus and system being measured.  This
decoherence process causes an increase in the disorder present in the
system, which is the intuition behind~(\ref{eq:msment_2}).  More
succinctly,~(\ref{eq:msment_2}) may be thought of as capturing the
notion that the total ensemble of possible quantum states is more
disordered after a measurement than it is before.

%
%
The importance of the static
constraints~(\ref{eq:maj_1_intro})-(\ref{eq:maj_2_intro}) and the
dynamics constraints~(\ref{eq:msment_1})-(\ref{eq:msment_2}) is
further reinforced by the fact that in each case there is a type of
converse to these equations.  In this introduction we focus only on
the more interesting case of the converse to the dynamic
constraints~(\ref{eq:msment_1}) and~(\ref{eq:msment_2}), however
rather similar remarks hold also for the static
constraints~(\ref{eq:maj_1_intro}) and~(\ref{eq:maj_2_intro}).
Suppose $p_i$ is a probability distribution, and $\rho$ and $\rho_i'$
are quantum states such that
\begin{eqnarray}
\lambda(\rho) & \prec & \sum_i p_i \lambda\left( \rho_i'
\right).
\end{eqnarray}
Then we will show that there exists a quantum measurement whose
measurement outcomes may be labelled by a {\em pair} of indices
$(i,j)$, such that for any fixed $i$ and for all $j$ the posterior
state of the quantum system after measurement is $\rho_i'$, and the
probabilities $p_{ij}$ for the $(i,j)$th measurement outcome satisfy
$\sum_j p_{ij} = p_i$.  Unfortunately, this result is not a tight
converse to equations~(\ref{eq:msment_1}) and~(\ref{eq:msment_2}), due
to the introduction of the extra index $j$, however for many purposes
it is a sufficiently strong converse.  We will show that even the
equations~(\ref{eq:msment_1}) and~(\ref{eq:msment_2}) together do not
completely characterize the quantum measurement process, however I
believe it likely that there is a simple characterization of the
measurement process along similar lines that may be expressed entirely
in terms of the eigenvalues of the prior and posterior states, and the
probabilities of the different measurement outcomes.  Of course, it is
true that the quantum measurement formalism already provides such a
characterization, in the form of a matrix equation, however equations
such as~(\ref{eq:msment_1}) and~(\ref{eq:msment_2}) provide far more
explicit information, and as such, are likely to be more useful in
practice.  We will demonstrate the utility of this approach by
application to the problem of entanglement transformation, simplifying
the proofs of several known results about entanglement
transformation\cite{Nielsen99a,Vidal99a,Vidal00b,Jonathan99a,Hardy99a}.

%
%
There is a striking level of symmetry in the
equations~(\ref{eq:maj_1_intro})-(\ref{eq:maj_2_intro}),
(\ref{eq:msment_1})-(\ref{eq:msment_2}), which we will also see in the
partial converse results.  It is obviously tempting to suggest that
this reflects some deeper underlying principle, much as Maxwell's
equations may be derived from a deeper action principle based on the
Faraday tensor, or the still deeper principles of gauge invariance and
relativity.  Unfortunately, I have not yet succeeding in obtaining a
satisfactory form for such a deeper principle.  Presumably, such a
deeper principle might assist in tightening the partial converse
results, or perhaps tightening the partial converses may shed light on
the origin of Equations~(\ref{eq:maj_1_intro})-(\ref{eq:maj_2_intro}),
(\ref{eq:msment_1})-(\ref{eq:msment_2}).

%
%
In explaining the intuitive meanings of the
equations~(\ref{eq:maj_1_intro})-(\ref{eq:maj_2_intro})
and~(\ref{eq:msment_1})-(\ref{eq:msment_2}) we have used language such
as the ``disorder'' present in a quantum state.  One might wonder if
it is possible to write down entropic statements capturing these
intuitions.  We will show that each of these equations in fact implies
an entropic statement whose content corresponds to the intuition we
have described.  Of course, entropic statements should really only be
interpreted in the aymptotic limit where we have a large number of
identical copies of a system available; the advantage of
Equations~(\ref{eq:maj_1_intro})-(\ref{eq:maj_2_intro})
and~(\ref{eq:msment_1})-(\ref{eq:msment_2}) is that they are stronger
forms of these asymptotic statements which may be applied to single
quantum systems.

%
%
This paper contains six fundamental results (together with a number of
applications), expressed in the four constraint equations,
(\ref{eq:maj_1_intro})-(\ref{eq:maj_2_intro}),
(\ref{eq:msment_1})-(\ref{eq:msment_2}), and the partial converses
to~(\ref{eq:maj_1_intro})-(\ref{eq:maj_2_intro}) and
(\ref{eq:msment_1})-(\ref{eq:msment_2}).  We now review antecedents of
these results in the existing literature.
Equation~(\ref{eq:maj_1_intro}) is an elementary consequence of
classic results in the theory of majorization.
Equation~(\ref{eq:maj_2_intro}) follows as a corollary of work of
Uhlmann\cite{Uhlmann70a}, Ruskai (unpublished, 1993) and
Nielsen\cite{Nielsen00c} on the relationship between mixed states and
probability distributions.  Equations~(\ref{eq:msment_1})
and~(\ref{eq:msment_2}) are implicit in the work of
Vidal\cite{Vidal00b} on entanglement transformation, and the partial
converse to~(\ref{eq:msment_1})-(\ref{eq:msment_2}) is implicit in the
work of Jonathan and Plenio\cite{Jonathan99a} on entanglement
transformation, building on earlier work by Nielsen\cite{Nielsen99a}.
A proof of Equation~(\ref{eq:msment_1}) in the context of entanglement
transformation has also been previously obtained by Jonathan, Nielsen,
Schumacher and Vidal (unpublished, 1999).  There are several
advantages to the point of view taken in the present paper.  First,
measurement is in some sense a more fundamental process than
entanglement transformation, and Equations~(\ref{eq:msment_1})
and~(\ref{eq:msment_2}) highlight the fundamental connection between
measurement and majorization for the first time, incidentally
explaining why there is a connection between entanglement
transformation and majorization: it arises as a result of a deeper
connection between measurement and majorization.  Second, the proofs
in the present paper are novel, and have the advantage of proceeding
from a more unified point of view than earlier work.  As a result they
are, perhaps, more elegant and informative than earlier proofs,
especially the proof of the partial converse
to~(\ref{eq:msment_1})-(\ref{eq:msment_2}), which is a substantial
improvement of and extension to existing constructions.  Several other
items of related work are also worth pointing out.  There is a
substantial mathematical literature on the problem of characterizing
the properties of sums $A+B$ of Hermitian matrices $A$ and $B$, and
Fulton\cite{Fulton00a} has written a nice review of recent progress on
this problem, which is closely related to the problem of mixing of
density matrices.  Hardy\cite{Hardy99a} has introduced techniques in
the context of entanglement transformation that can be used to
prove~(\ref{eq:msment_1}) and the partial converse
to~(\ref{eq:msment_1})-(\ref{eq:msment_2}).  Fuchs and Jacobs
(unpublished, 2000) have obtained a beautiful and quite different
proof of~(\ref{eq:msment_1}), after hearing of the result from
Nielsen.  Finally, the procedure described in this paper to prove the
partial converse to~(\ref{eq:msment_1})-(\ref{eq:msment_2}) is a
generalization of the procedures for entanglement transformation for
pure states found by Nielsen in \cite{Nielsen99a}, and subsequently
improved in independent work by Hardy, Jonathan and Nielsen (described
in Chapter~12 of \cite{Nielsen00a}), by Jensen and
Schack\cite{Jensen00a}, and by Werner (unpublished, 2000).

%
%
The paper is structured as follows.  We begin in
Section~\ref{sec:background} by reviewing the two main tools that will
be used in this paper, the theory of generalized measurements in
quantum mechanics, and the mathematical theory of majorization.
Section~\ref{sec:main} contains proofs of the static
constraints~(\ref{eq:maj_1_intro}) and~(\ref{eq:maj_2_intro}) on the
mixing of quantum states, and the dynamic
constraints~(\ref{eq:msment_1}) and~(\ref{eq:msment_2}) on quantum
measurement, and explores some elementary consequences of these
results.  In Section~\ref{sec:converse} we prove the partial converses
to~(\ref{eq:maj_1_intro})-(\ref{eq:maj_2_intro})
and~(\ref{eq:msment_1})-(\ref{eq:msment_2}).
Section~\ref{sec:entanglement} explains how the results of the present
paper may be used to obtain simplified proofs of known results about
entanglement transformation.  Finally, Section~\ref{sec:conc}
concludes the paper with a discussion of some open problems and future
directions.

\section{Generalized measurements and majorization}
\label{sec:background}

Before proceeding to the main results of the paper it is useful to
first review some background material on generalized measurements and
the mathematical theory of majorization.  All discussion in this and
succeeding sections is to be understood in the context of
finite-dimensional vector spaces, although infinite-dimensional
modifications seem likely to hold, perhaps with some technical
modifications.

\subsection{Generalized measurements}
\label{subsec:gen_meas}

%
%
In this paper we use the {\em generalized measurements formalism} as
our basic tool for the description of quantum measurements.  The
theory of generalized quantum measurements is an extension of the
projective measurements described in most quantum mechanics textbooks.
The reason the generalized measurements formalism is adopted is
because it is bettter adapted to the description of many realistic
quantum measurement schemes.  However, it is important to appreciate
that the generalized measurement formalism follows from standard
quantum mechanics, in the sense that any generalized measurement can
be understood as arising from the combination of unitary evolution
and a projective measurement, a correspondence made explicit below.
Nevertheless, the formalism of generalized measurements is in many
ways more useful and mathematically elegant than the standard
formulation of quantum measurement in terms of projectors.  More
detailed introductions to the theory of generalized measurements may
be found in \cite{Kraus83a,Schumacher96a,Nielsen00a,Gardiner91a}.

%
%
Mathematically, a generalized measurement is specified by a set $\{
E_i \}$ of {\em measurement matrices} satisfying the {\em completeness
  relation} $\sum_i E_i^{\dagger} E_i = I$.  The index $i$ on the
measurement matrices is in one-to-one correspondence with the possible
outcomes that may occur in the measurement.  The rule used to connect
the measurement matrices to physics is that if the prior state of the
quantum system is $\rho$ then the outcome $i$ occurs with probability
$p_ i = \mbox{tr}(E_i \rho E_i^{\dagger})$, and the posterior state is
given by $\rho_i' = E_i \rho E_i^{\dagger} / \mbox{tr}(E_i \rho
E_i^{\dagger})$.

%
%
Generalized measurements are obviously more general than the
projective measurements described in most textbooks.  Projective
measurements have the feature that they are {\em repeatable}, in the
sense that if one performs a projective measurement twice in a row on
a quantum system, then one will obtain the same result both times.  By
contrast, most real measurements don't have this feature of being
repeatable, which tips us off to the need for the formalism of
generalized measurements.  Nevertheless, even the generalized
measurement formalism can be understood in terms of projective
measurements as follows: the effect of a generalized measurement on a
quantum system is {\em equivalent} to a unitary interaction between
the system being measured and another ``ancilla'' system, followed by
a projective measurement on the ancilla system.  More precisely,
suppose $\{ E_i \}$ is a set of measurement matrices satisfying the
completeness relation $\sum_i E_i^{\dagger} E_i = I$.  We introduce an
ancilla system with orthonormal basis elements $|i\rangle$ indexed by
the possible measurement outcomes.  Define a matrix $U$ acting on the
joint quantum system-ancilla by the action:
\begin{eqnarray}
  U|\psi\rangle |0\rangle \equiv \sum_i E_i |\psi\rangle |i\rangle,
\end{eqnarray}
where $|0\rangle$ is some standard state of the ancilla and
$|\psi\rangle$ is an arbitrary state of the quantum system being
measured.  It is easy to show using the completeness relation $\sum_i
E_i^{\dagger} E_i = I$ that $U$ can be extended to a unitary matrix
acting on the entire state space of the joint system.  Suppose we
perform the unitary transformation $U$ on the joint quantum
system-ancilla, and then do a projective measurement of the ancilla in
the $|i\rangle$ basis.  It is then easily checked that the result of
the measurement is $i$ with probability $p_i = \mbox{tr}(E_i \rho
E_i^{\dagger})$ and the corresponding post-measurement state of the
system is $\rho_i' = E_i \rho E_i^{\dagger} / \mbox{tr}(E_i \rho
E_i^{\dagger})$.  Thus, the effect on the quantum system is exactly as
we have described above for a generalized quantum measurement.
Conversely, it is not difficult to verify that the effect of a unitary
interaction between system and ancilla followed by a projective
measurement on the ancilla can always be understood in terms of a
generalized measurement (see for example Chapter 8 of~\cite{Nielsen00a}).

\subsection{Majorization}
\label{subsec:majorize}

Our primary tool in the study of mixing and measurement in quantum
mechanics is the theory of majorization, whose basic elements we now
review.  The following review only covers elementary aspects of the
theory of majorization, and the reader is referred to Chapters~2 and~3
of~\cite{Bhatia97a}, \cite{Marshall79a} or~\cite{Alberti82a} for more
extensive background.

%
%
The basic motivation for majorization is to capture what it means to
say that one probability distribution is ``more mixed'' than another.
Suppose $x = (x_1,\ldots,x_d)$ and $y=(y_1,\ldots,y_d)$ are two
$d$-dimensional real vectors; we usually suppose in addition that $x$
and $y$ are probability distributions, that is, the components are
non-negative and sum to one, but the following definitions apply in
the case of general $x$ and $y$ as well.  The relation $x \prec y$,
read ``$x$ is majorized by $y$'', is intended to capture the notion
that $x$ is more mixed (i.e. disordered) than $y$.  To make the formal
definition, we introduce the notation $\downarrow$ to denote the
components of a vector rearranged into non-increasing order, so
$x^{\downarrow} = (x_1^{\downarrow},\ldots,x_d^{\downarrow})$, where
$x_1^{\downarrow} \geq x_2^{\downarrow} \geq \ldots \geq
x_d^{\downarrow}$.  We say that $x$ is majorized by $y$ and write $x
\prec y$, if
\begin{eqnarray}
  \sum_{j=1}^k x_j^{\downarrow} \leq \sum_{j=1}^k y_j^{\downarrow},
\end{eqnarray}
for $k=1,\ldots,d-1$, and with the inequality holding with equality
when $k = d$.

%
%
It is perhaps not so clear how this definition connects with any
natural notion of comparative disorder.  We will state but not prove a
remarkable result connecting majorization to a natural notion of
mixing.  It can be shown (see Chapter~2 of \cite{Bhatia97a}) that $x
\prec y$ if and only if $x = \sum_i p_i P_i y$, where the $p_i$s form
a probability distribution and the $P_i$s are permutation matrices.
Thus, when $x \prec y$ we can imagine that $y$ is the input
probability distribution to a noisy channel which randomly permutes
the symbols sent through the channel, inducing an output probability
distribution $x$.  From this characterization many other important
results follow with minimal effort; for example, it can easily be
shown that if $x \prec y$ then the Shannon entropy of the distribution
$x$ must be at least as great as that of $y$.

%
%
The connection between majorization and quantum mechanics arises
primarily as a result of {\em Horn's lemma} (proved in \cite{Horn54a};
for a simple proof see \cite{Nielsen00c}), which states that $x \prec
y$ if and only if there exists a unitary matrix $u = (u_{ij})$ such
that $x_i = \sum_j |u_{ij}|^2 y_j$.  This fundamental relationship
between majorization and unitarity ensures many close connections
between majorization and quantum mechanics.

%
%
As an elementary consequence of Horn's lemma we have {\em Ky Fan's
  maximum principle}, which states that for any Hermitian matrix $A$,
the sum of the $k$ largest eigenvalues of $A$ is the maximum value of
$\mbox{tr}(AP)$, where the maximum is taken over all $k$-dimensional
projectors $P$,
\begin{eqnarray}
\sum_{j=1}^k \lambda_j(A) = \max_P \mbox{tr}(AP).  
\end{eqnarray}
To see this, note that choosing $P$ to be the projector onto the space
spanned by the $k$ eigenvectors of $A$ with the $k$ largest
eigenvalues results in $\mbox{tr}(AP) = \sum_{j=1}^k \lambda_j(A)$.
The proof of Ky Fan's maximum principle will be completed if we can
show that $\mbox{tr}(AP) \leq \sum_{j=1}^k \lambda_j(A)$ for any
$k$-dimensional projector $P$.  To see this, let
$|e_1\rangle,\ldots,|e_d\rangle$ be an orthonormal basis chosen such
that $P = \sum_{j=1}^k |e_k\rangle \langle e_k|$.  Let
$|f_1\rangle,\ldots,|f_d\rangle$ be an orthonormal set of eigenvectors
for $A$, ordered so the corresponding eigenvalues are in
non-increasing order.  Then
\begin{eqnarray}
  \langle e_j|A|e_j\rangle = \sum_{k=1}^d |u_{jk}|^2 \lambda_k(A),
\end{eqnarray}
where $u_{jk} \equiv \langle e_j|f_k\rangle$ is unitary.  By Horn's
lemma it follows that $(\langle e_j|A|e_j\rangle ) \prec \lambda(A)$,
which implies that
\begin{eqnarray}
  \mbox{tr}(AP) = \sum_{j=1}^k \langle e_j|A|e_j\rangle \leq \sum_{j=1}^k
  \lambda_j(A),
\end{eqnarray}
as required.

%
%
Ky Fan's maximum principle gives rise to a useful constraint on the
eigenvalues of a {\em sum} of two Hermitian matrices, that
$\lambda(A+B) \prec \lambda(A)+\lambda(B)$.  To see this, choose a
$k$-dimensional projector $P$ such that
\begin{eqnarray}
  \sum_{j=1}^k \lambda_j(A+B) & = & \mbox{tr}((A+B)P) \\
  & = & \mbox{tr}(AP)+\mbox{tr}(BP) \\
  & \leq & \sum_{j=1}^k \lambda_j(A) + \sum_{j=1}^k \lambda_j(B),
\end{eqnarray}
where the last line also follows from Ky Fan's maximum principle.

%
%
Another consequence of Horn's lemma is that given a density matrix
$\rho$ and a probability distribution $p_i$ there exist pure states
$|\psi_i\rangle$ such that $\rho = \sum_i p_i |\psi_i\rangle \langle
\psi_i|$ if and only if $(p_i) \prec \lambda(\rho)$ (see
\cite{Nielsen00c,Uhlmann70a}; this result was also obtained in
unpublished work by Ruskai (1993)), where it is understood that if the
vector $(p_i)$ contains more terms than the vector $\lambda(\rho)$
then the vector $\lambda(\rho)$ is to be ``padded'' with extra zero
terms.  The proof of this result is simply to combine Horn's lemma
with the classification of ensembles $\{ p_i, |\psi\rangle \}$
consistent with a given density matrix $\rho$, as discovered
independently by Schr\"odinger\cite{Schrodinger36a},
Jaynes\cite{Jaynes57b} and Hughston, Jozsa and
Wootters\cite{Hughston93a}.  See \cite{Nielsen00c} for the details of
the proof.

%
%
This notion of ``padding'' vectors of unequal dimension so they can be
compared by the majorization relation is surprisingly useful, and we
adopt the general convention that when $x$ and $y$ are of different
dimension then $x \prec y$ means that $\tilde x \prec \tilde y$, where
$\tilde x$ and $\tilde y$ are padded with extra zero components to
ensure that they have the same dimension.  For example, $(1/3,1/3,1/3)
\prec (1/2,1/2)$ since $(1/3,1/3,1/3) \prec (1/2,1/2,0)$.  It is easy
to check that this extended notion of majorization is well-defined,
provided $x$ and $y$ both have non-negative components, and this will
be the case for all the applications in this paper.  Similarly, it is
often useful to write $x = y$ provided the padded versions of $x$ and
$y$ are equal, that is, the non-zero entries of $x$ and $y$ are equal.
With these conventions, it is easy to see that algebraic manipulations
proceed exactly as one would expect.  For example, for non-negative
real vectors $w,x,y,z$ if $w \prec x, x = y, y \prec z$ then obviously
$w \prec z$, even if all four vectors have different dimensionality.
We occasionally make use of such elementary observations in proofs,
without explicit comment.

%
%
The final result about majorization we shall need is that if $P_i$ are
a set of orthogonal projectors such that $\sum_i P_i = I$, and $\rho$
is a density matrix, then \cite{Bhatia97a}
\begin{eqnarray} \label{eq:maj_proj}
  \lambda\left( \sum_i P_i \rho P_i \right) \prec \lambda(\rho).
\end{eqnarray}
Intuitively, if a projective measurement of a quantum system is
performed, but we do not learn the result of the measurement, then the
state of the system after measurement is more mixed than it was
before.  One way of proving this relation is via Horn's lemma; a
sketch follows.  First, note that it suffices to prove that $\lambda(P
\rho P +Q \rho Q) \prec \lambda(\rho)$, where $P$ and $Q = I-P$ are
two orthogonal projectors satisfying $P+Q = I$.  Once this is proved,
the general relation~(\ref{eq:maj_proj}) follows by a simple
induction.  However, if we define a unitary matrix $U \equiv P-Q$ then
it is easy to verify that
\begin{eqnarray}
  P \rho P +Q \rho Q = \frac{\rho+U \rho U^{\dagger}}{2}.
\end{eqnarray}
Applying Horn's lemma and the easily proved fact that if $x_1 \prec y$
and $x_2 \prec y$ then $(x_1+x_2)/2 \prec y$, it follows with a little
simple linear algebra that $\lambda(P \rho P +Q\rho Q) \prec
\lambda(\rho)$.

\section{Proof of constraints on mixing and measurement in quantum mechanics}
\label{sec:main}

%
%
In this section we prove the four constraints,
(\ref{eq:maj_1_intro})-(\ref{eq:maj_2_intro}),
(\ref{eq:msment_1})-(\ref{eq:msment_2}).  The first and second of
these are static constraints on the mixing of quantum states, proved
in Subsection~\ref{subsec:static}.  The third and fourth constraint
equations are dynamic constraints on the quantum measurement process,
proved in Subsection~\ref{subsec:dynamic}.  Finally, some simple
consequences of these results are dicussed in
Subsection~\ref{subsec:consequences}.

\subsection{Static constraints on mixing quantum states}
\label{subsec:static}

%
%
{\bf Theorem 1:} Suppose $\rho = \sum_i p_i \rho_i$ is a convex
combination of quantum states $\rho_i$ with probabilities $p_i$.  Then
\begin{eqnarray}
  \label{eq:static_1} 
  \lambda\left( \rho \right) & \prec & \sum_i p_i \lambda\left( \rho_i
  \right) \\
  \label{eq:static_2}
  \bigoplus_i p_i \lambda \left( \rho_i \right) & \prec & \lambda \left(
  \rho \right).
\end{eqnarray}

%
%
{\bf Proof of~(\ref{eq:static_1}):} This is an immediate consequence
of the fact that $\lambda(A+B) \prec \lambda(A)+\lambda(B)$ for any
two Hermitian matrices $A$ and $B$, as proved in
Subsection~\ref{subsec:majorize}.

%
%
{\bf Proof of~(\ref{eq:static_2}):} As noted in
Subsection~\ref{subsec:majorize}, if a density matrix $\rho$ can be
written as a convex combination of pure states $|\psi_i\rangle$, $\rho
= \sum_i p_i |\psi_i\rangle \langle \psi_i|$, then it follows that
$(p_i) \prec \lambda(\rho)$, where $(p_i)$ denotes the vector whose
entries are the probabilities $p_i$.  Equation~(\ref{eq:static_2}) is
a corollary of this result.  To see this, note that if $r_{ij}$ are
the eigenvalues of $\rho_i$ and $|i,j\rangle$ the corresponding
orthonormal eigenvectors then~(\ref{eq:static_2}) is equivalent to the
equation
\begin{eqnarray}
  (p_i r_{ij}) \prec \lambda(\rho),
\end{eqnarray}
which follows from the results of Subsection~\ref{subsec:majorize} and
the observation that
\begin{eqnarray}
\rho = \sum_i p_i \rho_i = \sum_{ij} p_i r_{ij} |i,j\rangle \langle i,j|.
\end{eqnarray}
This completes the proof of Theorem~1. \hfill\rule{2.4mm}{2.4mm}

\subsection{Dynamical constraints on quantum measurement}
\label{subsec:dynamic}

%
%
{\bf Theorem 2:} Suppose $\{ E_i \}$ is a set of measurement matrices
satisfying the completeness relation $\sum_i E_i^{\dagger} E_i = I$.
Then the quantum measurement described by these matrices must
satisfy the following four constraints:
\begin{eqnarray}
  \label{eq:maj_1}
  \lambda\left( \sum_i E_i \rho E_i^{\dagger} \right) & \prec & \sum_i
  \lambda\left(E_i \rho E_i^{\dagger} \right) \\
  \label{eq:maj_2}
  \bigoplus_i \lambda\left(E_i \rho E_i^{\dagger} \right) & \prec &
  \lambda \left( \sum_i E_i \rho E_i^{\dagger} \right) \\
  \label{eq:maj_3}
  \lambda\left( \rho \right) & \prec & \sum_i \lambda \left( E_i \rho
  E_i^{\dagger} \right) \\
  \label{eq:maj_4}
  \bigoplus_i \lambda \left( E_i \rho E_i^{\dagger} \right) & \prec &
  \lambda\left( \rho \right).
\end{eqnarray}

%
%
A slightly different way of stating Theorem~2 is to define $p_i$ to be
the probability of obtaining outcome $i$ when the measurement defined
by the matrices $\{ E_i \}$ is performed on the system, and let
$\rho_i' = E_i \rho E_i^{\dagger}/\mbox{tr}(E_i \rho E_i^{\dagger})$
be the corresponding posterior states.  Then the following four
equations are equivalent to~(\ref{eq:maj_1})-(\ref{eq:maj_4}):
\begin{eqnarray} \label{eq:maj_1_intuit}
  \lambda\left( \sum_i p_i \rho_i' \right) & \prec & \sum_i
  p_i \lambda\left(\rho_i' \right) \\ \label{eq:maj_2_intuit}
  \bigoplus_i p_i \lambda\left(\rho_i' \right) & \prec &
  \lambda \left( \sum_i p_i \rho_i' \right) \\ \label{eq:maj_3_intuit}
  \lambda\left( \rho \right) & \prec & \sum_i p_i \lambda \left( \rho_i'
  \right) \\ \label{eq:maj_4_intuit}
  \bigoplus_i p_i \lambda \left( \rho_i' \right) & \prec &
  \lambda\left( \rho \right).
\end{eqnarray}

%
%
Theorem~2 is a fundamental constraint on the dynamics that may occur
during a quantum measurement.  Equations~(\ref{eq:maj_1_intuit})
and~(\ref{eq:maj_2_intuit}) are, of course, merely the dynamical
expression of the static constraints found earlier in Theorem~1.
Equations~(\ref{eq:maj_3_intuit}) and~(\ref{eq:maj_4_intuit})
represent novel constraints of an essentially dynamical nature,
connecting as they do the prior and posterior states of the quantum
measurement.  Intuitively, Equation~(\ref{eq:maj_3_intuit}) captures
the notion that a quantum measurement ``gains information'' (on
average) about a quantum state, since it says that the eigenvalues of
the initial state $\rho$ are, on average, more disordered than the
eigenvalues of the posterior states $\rho_i'$.  Intuitively, the
second dynamic constraint,~(\ref{eq:maj_4_intuit}) captures the notion
that the {\em total ensemble} of possible quantum states is more
disordered after the measurement than before.
Thus,~(\ref{eq:maj_3_intuit}) and~(\ref{eq:maj_4_intuit}) represent
complementary constraints on the evolution of a quantum system during
a quantum measurement process.

%
%
The constraints~(\ref{eq:maj_1_intuit})-(\ref{eq:maj_4_intuit}) are
applicable even for very complex measurement processes.  For example,
a single mode cavity undergoing direct photodetection by an ideal
photodetector can be described by a special case of the generalized
measurements formalism known as the {\em quantum trajectories} or {\em
  stochastic Schr\"odinger equation} picture
(see~\cite{Plenio98a,Zoller97a} for a review and references).  In this
picture, if the system is started in the state $\rho$ then the final
state of the system is $\rho_h$, where ``$h$'' is used here to denote
not just a single measurement outcome, but rather the complete history
recorded by the photodetector, that is, all the times at which
photocounts occurred.  Then~(\ref{eq:maj_3_intuit})
and~(\ref{eq:maj_4_intuit}) may be written as
\begin{eqnarray}
  \lambda(\rho) & \prec & \int d\mu(h) \lambda(\rho_h) \\
  \bigoplus_h d\mu(h) \lambda(\rho_h) & \prec & \lambda(\rho),
\end{eqnarray}
where the integral is a functional integral over all possible
photodetection histories, and $d\mu(h)$ is the corresponding measure
on histories.

%
%
{\bf Proof of Theorem~2:} The first two equations of
Theorem~2,~(\ref{eq:maj_1}) and~(\ref{eq:maj_2}), are immediate
consequences of the deeper {\em static} constraints on quantum
mechanics introduced in Theorem~1; here we are merely enumerating the
implications these static constraints have for dynamics.  The
remaining constraints,~(\ref{eq:maj_3}) and~(\ref{eq:maj_4}), are
genuine quantum dynamical constraints relating the prior and posterior
states of a quantum measurement.

%
%
{\bf Proof of~(\ref{eq:maj_3}):} Suppose $\rho$ is a positive matrix
which can be written in the block form:
\begin{eqnarray}
  \rho = \left[ 
        \begin{array}{cc} 
        A & X \\ 
        X^{\dagger} & B
      \end{array} \right].
\end{eqnarray}
For our purposes $\rho$ will most often be a density matrix (and thus
satisfy $\mbox{tr}(\rho) = 1$), but the results we prove hold for a
general positive matrix.  We will show that $\lambda(\rho) \prec
\lambda(A)+\lambda(B)$.  (Recall our conventions on padding, which
imply that the vectors of eigenvalues for $A$ and $B$ are to be
extended by zeroes in such a way that they contain as many entries as
the vector of eigenvalues of $\rho$).  $\rho$ is a positive matrix, so
there must exist a matrix $D = [ D_1 | D_2 ]$ such that $\rho =
D^{\dagger} D$, where the matrices $D_1$ and $D_2$ have the same
number of columns as $A$ and $B$, respectively, and both have the same
number of rows as $\rho$.  Thus we have
\begin{eqnarray}
  \left[ \begin{array}{cc} 
      A & X \\ X^{\dagger} & B
      \end{array} \right]
    = D^{\dagger} D =   \left[ \begin{array}{cc} 
      D_1^{\dagger} D_1 & D_1^{\dagger} D_2 \\ 
      D_2^{\dagger} D_1 & D_2^{\dagger} D_2
      \end{array} \right],
\end{eqnarray}
from which we read off $A = D_1^{\dagger} D_1$ and $B = D_2^{\dagger}
D_2$.  Using the results of Subsection~\ref{subsec:majorize} and the
fact that the eigenvalues of a product $EF$ of matrices $E$ and $F$
are the same as the eigenvalues of $FE$, up to padding by zeroes, we
see that
\begin{eqnarray}
  \lambda(\rho) & = & \lambda(D^{\dagger} D) \\
  & = & \lambda(D D^{\dagger}) \\
  & = & \lambda(D_1 D_1^{\dagger}+D_2D_2^{\dagger}) \\
  & \prec & \lambda(D_1 D_1^{\dagger})+\lambda(D_2D_2^{\dagger}) \\
  & = & \lambda(D_1^{\dagger} D_1)+\lambda(D_2^{\dagger}D_2) \\
  & = & \lambda(A)+\lambda(B),
\end{eqnarray}
and thus $\lambda(\rho) \prec \lambda(A)+\lambda(B)$, as claimed.
This method for eliminating off-diagonal block terms was introduced by
Wielandt to connect the Weyl and Aronszajn inequalities (cited
as~\cite{Wielandt67a} in Chapter~3 of~\cite{Bhatia97a}.)

%
%
As a straightforward consequence we see by induction that for any
positive matrix $\rho$ and complete set of orthogonal projectors $\{
P_i \}$:
\begin{eqnarray} \label{eq:Weyl_gen}
  \lambda(\rho) \prec \sum_i \lambda(P_i \rho P_i)
\end{eqnarray}
Extending even further, suppose $\{ E_i \}$ is any set of measurement
matrices defining a generalized measurement, and $\rho$ is a positive
matrix.  As in Subsection~\ref{subsec:gen_meas} we can introduce an
ancilla system with an orthonormal basis $|i\rangle$ in one-to-one
correspondence with the indices on the measurement matrices $E_i$ and
define a unitary matrix $U$ which has the action
\begin{eqnarray}
  U|\psi\rangle |0\rangle = \sum_i E_i |\psi\rangle |i\rangle,
\end{eqnarray}
where $|0\rangle$ is some standard state of the ancilla.  Then we have
$\lambda(\rho) = \lambda(\rho \otimes |0\rangle \langle 0|)$,
since the non-zero eigenvalues of $\rho$ and $\rho \otimes |0\rangle
\langle 0|$ are the same.  Simple algebra and~(\ref{eq:Weyl_gen})
imply that
\begin{eqnarray}
\lambda(\rho) & = & \lambda(U (\rho \otimes |0\rangle \langle 0|)U^{\dagger}) \\
  & \prec & \sum_i \lambda( (I\otimes |i\rangle\langle i|) U
  (\rho \otimes |0\rangle \langle 0|) U^{\dagger} (I \otimes |i\rangle\langle i|))  \\
  & = & \sum_i \lambda( E_i \rho E_i^{\dagger} \otimes |i\rangle \langle i|) \\
  & = & \sum_i \lambda(E_i \rho E_i^{\dagger}),
\end{eqnarray}
where in the last line we used $\lambda(E_i \rho E_i^{\dagger} \otimes
|i\rangle \langle i|) = \lambda(E_i \rho E_i^{\dagger})$, since the
non-zero entries agree.  This completes the proof of~(\ref{eq:maj_3}).

%
%
{\bf Proof of~(\ref{eq:maj_4}):} Again, let $U$ be the unitary matrix
constructed in Subsection~\ref{subsec:gen_meas} to implement the
measurement described by the measurement matrices $\{ E_i \}$, namely,
any unitary matrix having the action
\begin{eqnarray}
  U|\psi\rangle|0\rangle = \sum_i E_i |\psi\rangle |i\rangle.
\end{eqnarray}
Again, we have $\lambda(\rho) = \lambda(\rho \otimes |0\rangle \langle
0|)$, since the non-zero eigenvalues of $\rho$ are the same as those
of $\rho \otimes |0\rangle \langle 0|$, and thus $\lambda(\rho) =
\lambda\left( U(\rho \otimes |0\rangle \langle 0| )
  U^{\dagger}\right)$.  It follows from Equation~(\ref{eq:maj_proj})
that
\begin{eqnarray}
  \lambda\left( \sum_i (I \otimes |i\rangle \langle i|) U(\rho \otimes |0\rangle
  \langle 0| ) U^{\dagger} (I \otimes |i\rangle \langle i|) \right) \prec \lambda(\rho),
\end{eqnarray}
and thus
\begin{eqnarray}
  \lambda\left( \sum_i E_i \rho E_i^{\dagger} \otimes |i\rangle \langle i|
  \right) \prec \lambda(\rho).
\end{eqnarray}
This last equation is obviously equivalent to the statement we set out
to prove,
\begin{eqnarray}
  \bigoplus_i \lambda\left( E_i \rho E_i^{\dagger} \right) \prec \lambda(\rho),
\end{eqnarray}
which concludes the proof of Theorem~2. \hfill\rule{2.4mm}{2.4mm}

\subsection{Consequences of the constraint equations}
\label{subsec:consequences}

%
%
The constraints proved in Theorems~1 and~2 are very strong and, not
surprisingly, have many interesting consequences.  We now elucidate a
few of these consequences using the notions of {\em Schur-concavity}
and {\em Schur-convexity}.  A Schur-convex function $f(\cdot)$ is a
real-valued function which preserves the majorization relation, in the
sense that if $x \prec y$ then $f(x) \leq f(y)$.  Simple necessary and
sufficient conditions for a function to be Schur-convex are known
\cite{Bhatia97a}, and many interesting functions are Schur-convex.
These include, for example, the function $x \rightarrow f(x) \equiv
\sum_{j=1}^d x_j^k$, for any $k \geq 1$.  Similarly, a {\em
  Schur-concave} function $f(\cdot)$ is one such that if $x \prec y$
then $f(x) \geq f(y)$.  Equivalently, $f(\cdot)$ is Schur-concave if
$-f(\cdot)$ is Schur-convex.  Perhaps the canonical example of a
Schur-concave function is the Shannon entropy $H(x) = -\sum_j x_j
\log_2(x_j)$, so that whenever $x \prec y$ it follows that $H(x) \geq
H(y)$, giving further justification to the intuitive notion that $x
\prec y$ means that $x$ is more disordered than $y$.  Applying the
Schur-concavity of Shannon's entropy to the results of Theorems~1
and~2 we obtain an attractive suite of results.  First, applying the
Schur-concavity of $H(\cdot)$ to~(\ref{eq:static_1}) gives
\begin{eqnarray}
  S(\rho) & \geq & H\left( \sum_i p_i \lambda(\rho_i) \right).
\end{eqnarray}
Applying the concavity of the Shannon entropy to the right hand side,
we obtain as a corollary the concavity of the von~Neumann entropy,
\begin{eqnarray}
  S(\rho) & \geq & \sum_i p_i S(\rho_i).
\end{eqnarray}
Applying the Schur-concavity of $H(\cdot)$ to~(\ref{eq:static_2}) and
doing some simple algebra gives
\begin{eqnarray}
  \sum_i p_i S(\rho_i) + H(p_i) \geq S(\rho).
\end{eqnarray}
This result was obtained previously by Lanford and
Robinson\cite{Lanford68a} using different techniques.  Applying the
Schur-concavity of $H(\cdot)$ to~(\ref{eq:maj_3_intuit}), followed by the
concavity of the Shannon entropy, gives
\begin{eqnarray}
  S(\rho) \geq \sum_i p_i S(\rho_i').
\end{eqnarray}
Essentially the same result has been obtained previously in the
context of entanglement transformation \cite{Bennett96a}, where it
expresses the fact that local processes cannot increase the amount of
entanglement present in a system.  Finally, applying the
Schur-concavity of $H(\cdot)$ to~(\ref{eq:maj_4_intuit}) gives the
beautiful inequality
\begin{eqnarray}
  H(p_i)+\sum_i p_i S(\rho_i') \geq S(\rho),
\end{eqnarray}
which implies that in order to lower the entropy of a system by an
amount $\Delta$, on average, the information $H(p_i)$ collected by the
measurement must be at least as large as $\Delta$.  This fact can be
seen as a quantum mechanical expression of the principle, expressed by
Landauer\cite{Landauer61a} and fleshed out by Bennett\cite{Bennett82b}
and Zurek\cite{Zurek89b}, that measurement of a physical system
carries with it a thermodynamic cost when the measurement record is
erased, and proper accounting of this cost enables one to solve the
conundrum posed by Maxwell's demon.  (See \cite{Bennett87a} for a
review.)

%
%
Applying the Schur-convexity of the functions $f(x) = \sum_i x_i^k$
for $k \geq 1$ to the results of Theorems~1 and~2 also give a number
of interesting constraints.  The arguments used are analogous to those
given above for the Shannon entropy, so the details will be omitted,
and we merely state the results:
\begin{eqnarray}
  \sum_i p_i^k \mbox{tr}\left(\rho_i^k\right) & 
  \leq \mbox{tr}\left(\rho^k\right) \leq & \sum_i p_i
  \mbox{tr}\left(\rho_i^k\right) \\
\sum_i p_i^k \mbox{tr}\left( (\rho_i')^k \right) & 
  \leq \mbox{tr}\left(\rho^k
  \right) \leq & \sum_i p_i \mbox{tr}\left( (\rho_i')^k \right).
\end{eqnarray}

\section{Partial converses to the constraints on mixing and
  measurement}
\label{sec:converse}

%
%
Given the constraints on mixing and measurement described in
Theorems~1 and~2 it is natural to ask if these constraints completely
{\em characterize} the processes of mixing and measurement,
respectively.  We will show below that the answer to this question is
{\em no}.  However, partial progress towards achieving simple
characterizations of mixing and measurement may be reported in the
form of a partial converse to Theorem~1, described below in
Subsection~\ref{subsec:con_mix}, and a partial converse to Theorem~2,
described in Subsection~\ref{subsec:con_meas}.

\subsection{Partial converse to the constraints on mixing}
\label{subsec:con_mix}

%
%
Given the constraints Theorem~1 imposes on mixing it is natural to ask
whether these constraints completely characterize the mixing process.
That is, given a density matrix $\rho$, probabilities $p_i$ and
vectors $\lambda_i$ with non-negative, non-increasing components which
sum to one, and such that
\begin{eqnarray}
  \label{eq:maj_1_rehash}
  \lambda(\rho) & \prec & \sum_i p_i \lambda_i \\ 
  \label{eq:maj_2_rehash}
  \bigoplus_i p_i \lambda_i & \prec & \lambda(\rho),
\end{eqnarray}
does it follow that there exist density matrices $\rho_i$ such that
$\lambda(\rho_i) = \lambda_i$ and $\rho = \sum_i p_i \rho_i$?

%
%
We will show below that the answer to this question is no, however I
suspect that some characterization along similar lines is possible.
Progress towards such a characterization can be reported in the form
of a partial converse to Theorem~1, which states that
provided~(\ref{eq:maj_1_rehash}) holds then there exist states
$\rho_{ij}$ and a probability distribution $p_{ij}$ such that
$\lambda(\rho_{ij}) = \lambda_i$, independent of the value of the
index $j$, and $p_i = \sum_j p_{ij}$ for each $i$, as well as $\rho =
\sum_{ij} p_{ij} \rho_{ij}$.  That is, in order to obtain a converse
to~(\ref{eq:maj_1_rehash}) we need to introduce an extra index, $j$.
We will show below that it is necessary to introduce the extra index
if only~(\ref{eq:maj_1_rehash}) is assumed as a hypothesis for the
converse.  Let's state and prove the partial converse as Theorem~3.

%
%
{\bf Theorem 3:} Suppose $\rho$ is a density matrix and $\lambda_i$
are vectors with non-negative, non-increasing components summing to
one.  Suppose $p_i$ are probabilities such that
\begin{eqnarray}
  \label{eq:thm_3}
  \lambda(\rho) \prec \sum_i p_i \lambda_i.
\end{eqnarray}
Then there exist density matrices $\rho_{ij}$ and a probability
distribution $p_{ij}$ such that $p_i = \sum_j p_{ij}$,
$\lambda(\rho_{ij}) = \lambda_i$, and $\rho = \sum_{ij} p_{ij}
\rho_{ij}$.

%
%
To prove Theorem~3 we need the result stated in
Subsection~\ref{subsec:majorize} that $x \prec y$ if and only if there
exist probabilities $q_j$ and permutation matrices $P_j$ such that $x
= \sum_j q_j P_j y$.  Applying this result with the
assumption~(\ref{eq:thm_3}) we obtain
\begin{eqnarray}
  \label{eq:thm_3_inter}
  \lambda(\rho) = \sum_{ij} p_i q_j P_j \lambda_i.
\end{eqnarray}
Working in the basis in which $\rho$ is diagonal, and defining
$\Lambda_i$ to be the diagonal matrix with diagonal entries
$\lambda_i$, we may set $p_{ij} \equiv p_i q_j$ and $\rho_{ij} \equiv
P_j \Lambda_i P_j^{\dagger}$, obtaining $p_i = \sum_j p_{ij}$ and
$\lambda(\rho_{ij}) = \lambda_i$.  Finally, the equation $\rho =
\sum_{ij} p_{ij} \rho_{ij}$ follows immediately from these definition
and~(\ref{eq:thm_3_inter}), completing the proof.

%
%
What of a tight converse to Theorem~1?  It is easy to see that it is
not possible to obtain a tight converse to~(\ref{eq:maj_1_rehash})
alone, as follows.  Suppose we choose $\rho = I/2$ to be the
completely mixed state of a single qubit, and define a probability
distribution on just one outcome, the trivial distribution $p_1 = 1$,
with corresponding vector $\lambda_1 = (1,0)$.  Clearly,
$\lambda(\rho) \prec \sum_i p_i \lambda_i$, yet it is not possible to
find a state $\rho_1$ such that $\rho = p_1 \rho_1$ and
$\lambda(\rho_1) = \lambda_1$.  Thus, in this example, it is necessary
to introduce extra indices, just as was done in Theorem~3.  

%
%
Might it be that conditions~(\ref{eq:maj_1_rehash})
and~(\ref{eq:maj_2_rehash}) together completely characterize the
mixing process?  The following example, due to Julia~Kempe, shows that
this is not the case.  Suppose we consider a qubit system, and choose
$\rho = \mbox{diag}(5/12,7/12)$, $p_1 = p_2 = 1/2$, and $\lambda_1 =
(1,0), \lambda_2 = (1/2,1/2)$.  It is easy to verify that
conditions~(\ref{eq:maj_1_rehash}) and~(\ref{eq:maj_2_rehash}) are
satisfied with these choices.  Unfortunately, it is not possible to
find states $\rho_1$ and $\rho_2$ with vectors of eigenvalues
$\lambda_1$ and $\lambda_2$ such that $\rho = p_1 \rho_1 + p_2
\rho_2$, since with these choices for $\lambda_1$ and $\lambda_2$ it
follows that $\rho_1$ must be a pure state and $\rho_2 = I/2$ the
completely mixed state, so $p_1 \rho_1 + p_2 \rho_2$ has eigenvalues
$3/4$ and $1/4$, which are not equal to $5/12$ and $7/12$.  Despite
this example, I believe it likely that conditions along the lines
of~(\ref{eq:maj_1_rehash}) and~(\ref{eq:maj_2_rehash}) may be used to
completely characterize the process of mixing in quantum mechanics.

\subsection{Partial converse to the constraints on measurement}
\label{subsec:con_meas}

%
%
Given the constraints Theorem~2 imposes on the quantum measurement
process it is natural to ask whether these constraints completely
characterize the possible posterior states and probabilities which may
occur in such a measurement?  That is, supposing $\rho$ is a density
matrix, $p_i$ is a probability distribution, and $\rho_i'$ are density
matrices such that
\begin{eqnarray}
  \label{eq:maj_3_rehash}
  \lambda\left( \rho \right) & \prec & \sum_i p_i \lambda \left( \rho_i'
  \right) \\
  \label{eq:maj_4_rehash}
  \bigoplus_i p_i \lambda \left( \rho_i' \right) & \prec &
  \lambda\left( \rho \right),
\end{eqnarray}
does it follow that there exist measurement matrices $\{ E_i \}$
satisfying the completeness relation $\sum_i E_i^{\dagger} E_i = I$
and giving the states $\rho_i'$ as posterior states, with
probabilities $p_i$, when the measurement is performed on a system
initially prepared in the state $\rho$?

%
%
We will show below that the answer to this question is no, however I
suspect that some characterization along similar lines is possible.
Progress towards such a characterization can be reported in the form
of a partial converse to Theorem~2, which states that provided the
relation~(\ref{eq:maj_3_rehash}) holds, then there is a quantum
measurement described by measurement matrices $\{ E_{ij} \}$ such that
the corresponding posterior states $\rho_{ij}'$ satisfy $\rho_{ij}' =
\rho_i$ for every $j$, and the measurement probabilities $p_{ij}$
satisfy $\sum_j p_{ij} = p_i$.  Thus, in order to obtain a converse
to~(\ref{eq:maj_3_rehash}) we need to introduce an extra index, $j$,
just as we did earlier in the partial converse to Theorem~1.  Also
analogously to that case, we show below that it is necessary to
introduce the extra index with only~(\ref{eq:maj_3_rehash}) as
hypothesis for the converse.  Let's state and prove the partial
converse as Theorem~4.

%
%
{\bf Theorem 4:} Suppose $\rho$ is a density matrix with vector of
eigenvalues $\lambda$, and $\sigma_i$ are density matrices with
vectors of eigenvalues $\lambda_i$.  Suppose $p_i$ are probabilities
such that
\begin{eqnarray}
  \label{eq:converse}
  \lambda \prec \sum_i p_i \lambda_i
\end{eqnarray}
Then there exist matrices $\{ E_{ij} \}$ and a probability
distribution $p_{ij}$ such that 
\begin{eqnarray}
\sum_{ij} E_{ij}^{\dagger} E_{ij} & = & I \\
E_{ij} \rho E_{ij}^{\dagger} & = & p_{ij} \sigma_i \\
\sum_j p_{ij} & = & p_i.
\end{eqnarray}

%
%
To prove Theorem~4, we again use the result that $x \prec y$ if and
only if there exist probabilities $q_j$ and permutation matrices $P_j$
such that $x = \sum_j q_j P_j y$.  By assumption we have $\lambda
\prec \sum_i p_i \lambda_i$ and thus there exist permutation matrices
$P_j$ and probabilities $q_j$ such that
\begin{eqnarray} \label{eq:thm_2_inter}
  \lambda = \sum_{ij} p_i q_j P_j \lambda_i.
\end{eqnarray}
Without losss of generality we may assume that $\rho$ and $\sigma_i$
are all diagonal in the same basis, with non-increasing diagonal
entries, since if this is not the case then it is an easy matter to
prepend or append unitary matrices to the measurement matrices to
obtain the correct transformation.  With this convention, we define
matrices $E_{ij}$ by
\begin{eqnarray} \label{eq:meas_def}
  E_{ij} \sqrt{\rho} \equiv \sqrt{p_i q_j} \sqrt{\sigma_i} P_j^{\dagger}.
\end{eqnarray}
In order for $E_{ij}$ to be well-defined by this formula alone it is
necessary that $\rho$ be invertible.  If this is not the case then the
$E_{ij}$ are defined on the support of $\rho$ by the
formula~(\ref{eq:meas_def}), and to act as the zero operator on the
orthocomplement of the support of $\rho$.  It is convenient to let $P$
be the projector onto the support of $\rho$.  Note that we have
\begin{eqnarray}
  \sqrt{\rho} \left(
  \sum_{ij} E_{ij}^{\dagger} E_{ij} \right) \sqrt{\rho} & = & \sum_{ij}
  p_i q_j P_j \sigma_i P_j^{\dagger}.
\end{eqnarray}
Comparing with~(\ref{eq:thm_2_inter}) we see that the right-hand side
of the last equation is just $\rho$ and thus
\begin{eqnarray}
  \sqrt{\rho} \left( \sum_{ij} E_{ij}^{\dagger} E_{ij} \right)
  \sqrt{\rho} = \rho,
\end{eqnarray}
from which we deduce that $\sum_{ij} E_{ij}^{\dagger} E_{ij} = P$, the
projector onto the support of $\rho$.  Letting $Q \equiv I- P$ be the
projector onto the orthocomplement of the support, we can append an
additional measurement matrix $E_{00} \equiv Q$ to the collection
$E_{ij}$ to ensure that the completeness relation $\sum_{ij}
E_{ij}^{\dagger} E_{ij} = I$ is satisifed.  Furthermore, from the
definition~(\ref{eq:meas_def}) it follows that
\begin{eqnarray}
  E_{ij} \rho E_{ij}^{\dagger} & = & p_i q_j \sigma_i,  
\end{eqnarray}
and thus upon performing a measurement defined by the measurement
matrices $\{ E_{ij} \}$ the result $(i,j)$ occurs with probability
$p_{ij} = p_i q_j$, $\sum_j p_{ij} = p_i$, and the post-measurement
state is $\sigma_i$.  This completes the proof of Theorem~4.

%
%
Theorem~4 is not a sharp converse to the condition of
Equation~(\ref{eq:maj_3_rehash}) because of the extra index $j$.
Introducing some such index is certainly necessary with the present
hypotheses, as may be seen by considering an example with $\lambda =
(1/2,1/2)$, and the trivial probability distribution on one outcome,
$p_1 = 1$, with $\lambda_1 = (1,0)$.  Then $\lambda \prec p_1
\lambda_1$, but it is clear that there does not exist an $E_1$ such
that $E_1\rho E_1^{\dagger} = \rho_1$, where $\lambda(\rho) = \lambda,
\lambda(\rho_1) = \lambda_1$ and $E_1^{\dagger}E_1 = I$, because the
last equation implies that $E_1$ must be unitary.  It is not difficult
to construct more complex examples to convince oneself that this
behaviour is generic.

%
%
Might it be that the conditions~(\ref{eq:maj_3_rehash})
and~(\ref{eq:maj_4_rehash}) together characterize the posterior states
and probabilities achievable through a quantum measurement?  The
following argument, due to Julia~Kempe and the author, shows that this
is not the case.  Suppose we consider a qubit system, and choose $\rho
= \mbox{diag}(5/12,7/12)$, $p_1 = p_2 = 1/2$, and $\rho_1' =
\mbox{diag}(1,0), \rho_2' = \mbox{diag}(1/2,1/2)$.  It is easy to
verify that conditions~(\ref{eq:maj_3_rehash})
and~(\ref{eq:maj_4_rehash}) are satisfied with these choices.
Unfortunately, it is not possible to find measurement matrices $E_1$
and $E_2$ satisfying $\sum_i E_i^{\dagger} E_i = I$ and giving
posterior states $\rho_1'$ and $\rho_2'$ with equal probabilites
$1/2$, when the state $\rho$ is measured.  This can be seen in a
variety of ways.  A simple direct way is to note that the purity of
$\rho_1'$ implies that $E_1$ must have the form $E_1 = \alpha
|a\rangle \langle b|$ for normalized states $|a\rangle$ and
$|b\rangle$, and some $\alpha > 0$.  Thus
\begin{eqnarray}
  E_2^{\dagger} E_2 & = & I-E_1^{\dagger} E_1 \\
  & = & I-\alpha^2 |b\rangle  \langle b|  \\
  & = & (1-\alpha^2) |b\rangle \langle b| +|c\rangle \langle c|,
\end{eqnarray}
where $|c\rangle$ is orthonormal to $|b\rangle$.  The polar
decomposition gives $E_2 = U \sqrt{E_2^{\dagger} E_2}$ for some
unitary $U$, so
\begin{eqnarray}
  E_2 = \sqrt{1-\alpha^2} U|b\rangle \langle b| + U|c\rangle \langle c|.
\end{eqnarray}
We are requring that $E_2 \rho E_2^{\dagger} = I/4$, so it must be the
case that $E_2$ is non-singular, and thus $\alpha < 1$.
Premultiplying by $E_2^{-1}$ and postmultiplying by
$(E_2^{\dagger})^{-1}$ gives
\begin{eqnarray}
  \rho = \frac{1}{4(1-\alpha^2)} |b\rangle \langle b| + \frac{1}{4}
  |c\rangle \langle c|.
\end{eqnarray}
Since $|b\rangle$ and $|c\rangle$ are orthonormal it follows that such
a $\rho$ cannot be equal to $\mbox{diag}(5/12,7/12)$, which is the
desired contradiction.  Despite this example, I believe it likely that
conditions along the lines of~(\ref{eq:maj_3_rehash})
and~(\ref{eq:maj_4_rehash}) may be used to characterize the process of
measurement in quantum mechanics.

\section{Entanglement transformation}
\label{sec:entanglement}

The problem of {\em entanglement transformation} is a natural context
in which the results of the present paper may be applied.  The problem
of entanglement transformation arises as a consequence of the
fundamental question of how may we convert one type of physical
resource into another, and there has been considerable effort devoted
to determining when it is possible to convert one type of entanglement
to another.  In~\cite{Nielsen99a} a connection was noted between
entanglement transformation and majorization, namely, that if
$|\psi\rangle$ and $|\phi\rangle$ are pure states of a bipartite
quantum system with components belonging to Alice ($A$) and Bob ($B$)
respectively, then Alice and Bob can transform the state
$|\psi\rangle$ into the state $|\phi\rangle$ using local operations on
their respective systems and classical communication between Alice and
Bob, if and only if
\begin{eqnarray}
  \label{eq:Nielsen}
  \lambda_{\psi} \prec \lambda_{\phi},
\end{eqnarray}
where $\lambda_{\psi}$ (respectively $\lambda_{\phi}$) is the vector
of eigenvalues of the reduced density matrix for Alice's system when
the joint system is in the state $|\psi\rangle$ ($|\phi\rangle$).  As
per usual, the components of such vectors are ordered into
non-increasing order.  This result has subsequently been generalized
by Vidal\cite{Vidal99a} to the case of conclusive transformation, and
even further by Jonathan and Plenio\cite{Jonathan99a} to the problem
where Alice and Bob are supplied with a state $|\psi\rangle$ and wish
to tranform this state into an {\em ensemble} of states in which the
state $|\phi_i\rangle$ occurs with probability $p_i$.  (See also
Hardy\cite{Hardy99a} for an instructive alternative approach to
results of this type.)  The necessary and sufficient condition for
such a transformation to be possible is that\cite{Jonathan99a}:
\begin{eqnarray}
  \label{eq:Jonathan}
  \lambda_{\psi} \prec \sum_i p_i \lambda_{\phi_i}.
\end{eqnarray}
We now explain how this result can be seen as an easy consequence of
the results proved in the present paper, and thus the connection
between majorization and entanglement is really a consequence of a
deeper connection between majorization and measurement.  By a result
of Lo and Popescu\cite{Lo97b}, it is possible to transform
$|\psi\rangle$ into the ensemble $\{ p_i, |\phi_i\rangle \}$ by local
operations and classical communication if and only if it is possible
to make the transformation via the following simplified procedure:
first, Alice performs a generalized measurement on her state, then
sends the result to Bob, who performs a unitary operation on his
system conditional on the outcome of the measurement Alice made.  Let
$\rho =\mbox{tr}_B(|\psi\rangle \langle \psi|)$ be the initial state
of Alice's system, and suppose Alice performs a quantum measurement
described by measurement matrices $E_i$, so that outcome $i$ occurs
with probability $p_i$ and $(E_i \otimes U_i) |\psi\rangle =
\sqrt{p_i} |\phi_i\rangle$, for some unitary operator $U_i$ acting on
Bob's system.  Considering Alice's system alone and observing that
that $E_i \rho E_i^{\dagger} = \sigma_i$, where $\sigma_i = p_i
\mbox{tr}(|\phi_i\rangle \langle \phi_i|)$, we deduce from Theorem~2
that
\begin{eqnarray}
  \lambda_{\rho} \prec \sum_i p_i \lambda_{\sigma_i},
\end{eqnarray}
which is equivalent to~(\ref{eq:Jonathan}).  To prove the converse,
suppose~(\ref{eq:Jonathan}) holds.  Then by Theorem~4 there exists a
quantum measurement described by measurement matrices $E_{ij}$, and
probabilities $p_{ij}$ such that
\begin{eqnarray}
  E_{ij} \rho E_{ij}^{\dagger} = p_{ij} \sigma_i; \,\,\,\, \sum_j p_{ij}
  = p_i.
\end{eqnarray}
The procedure for Alice and Bob to produce the ensemble is for Alice
to perform the measurement described by the set $E_{ij}$.  The
post-measurement state $|\phi_{ij}\rangle$ is then a purification
\cite{Nielsen00a} of the state $\sigma_i$, and it can be shown (see
\cite{Hughston93a} or Section~2.5 of~\cite{Nielsen00a}) that by
performing an appropriate unitary transformation Bob can convert the
state $|\phi_{ij}\rangle$ into the state $|\phi_i\rangle$, with total
probability $p_i$ of obtaining the state $|\phi_i\rangle$.  Thus
Equation~(\ref{eq:Jonathan}) represents a necessary and sufficient
condition for it to be possible to transform the state $|\psi\rangle$
into the ensemble $\{ p_i,|\phi_i\rangle \}$ by local operations and
classical communication.

\section{Conclusion}
\label{sec:conc}

%
%
We have shown that there are strong fundamental constraints on the
processes of mixing and measurement in quantum mechanics that may be
naturally expressed in the language of majorization.  Although the
results in the present paper don't completely characterize these
processes, they suggest that there may exist a simple set of
conditions which substantially simplify the usual characterization of
these processes via operator equations.  Another interesting direction
for further research is to generalize the constraints on measurements
obtained in this paper to better understand how two or more states may
transform simultaneously under a measurement.  Once again, although
this problem is in principle already ``solved'', in the sense that
there is an operator equation specifying exactly what transformations
may occur, results such as those in the present paper and in
\cite{Chefles99a} indicate that much more explicit characterizations
may be possible.  Such explicit conditions are likely to have
applications to fundamental problems such as the problem of
transformation of mixed state entanglement\cite{Bennett96a}, and to
the problem of determining to what extent the acquisition of
information about the identity of a quantum state disturbs the system
being measured\cite{Fuchs96b}.

\section*{acknowledgments}
Thanks to Dave~Beckman, Sumit~Daftuar, Julia~Kempe, Beth~Ruskai and
Armin~Uhlmann for discussions about majorization.  Thanks also to
Manny~Knill for encouragement and bringing reference \cite{Fulton00a}
to my attention, and to Chris~Fuchs, Kurt~Jacobs and Reinhard~Werner
for bringing their unpublished work to my attention.  This work was
supported by a Tolman Fellowship, and by DARPA through the Quantum
Information and Computing Institute (QUIC) administered through the
ARO.


\end{multicols}

\end{document}